\documentclass[prl,amsmath,amssymb]{revtex4}
\usepackage{graphicx}
\usepackage{dcolumn}
\usepackage{bm}
\begin{document}
\title{Criticality in correlated quantum matter}
\author{Angela Kopp and Sudip Chakravarty$^{*}$}
\affiliation{Department of Physics and Astronomy, University of
California Los Angeles, Los Angeles, CA 90095-1547}
\date{\today}
\begin{abstract}
{\bf  At quantum critical points (QCP) \cite{Pfeuty:1971,Young:1975,Hertz:1976,Chakravarty:1989,Millis:1993,Chubukov:1994,Coleman:2005} there are quantum fluctuations on all length scales,  from microscopic to macroscopic lengths, which, remarkably, can be observed at finite temperatures, the regime to which all experiments are necessarily confined.  A fundamental question is how high in temperature can the effects of quantum criticality persist? That is, can physical observables be described in terms of universal scaling functions originating from the QCPs?  Here we answer these questions by examining exact solutions of models of correlated systems and find that the temperature can be surprisingly high. As a powerful illustration of quantum criticality, we  predict  that the zero temperature superfluid density, $\rho_{s}(0)$, and the transition temperature, $T_{c}$, of the cuprates are related by $T_{c}\propto\rho_{s}(0)^y$, where the exponent $y$ is different at the two edges of the superconducting dome, signifying the respective QCPs. This relationship can be tested in high quality crystals. }
\end{abstract}
\email[Correspondence and requests for materials should be addressed to S.C.  ] { sudip@physics.ucla.edu}

\maketitle

 Do quantum critical points (QCPs) provide a powerful framework for understanding complex correlated many body problems? Do they shed new light on quantum mechanics of macroscopic systems, providing, for example, a deeper understanding of entanglement? Answers to many such questions require a precise recognition of the experimentally observable regime of quantum criticality.  
 An important class of QCPs  are analogs of classical critical points, but in a dimension higher than the actual spatial dimension of the system; that is, a quantum critical point in $d$ spatial dimensions is equivalent to a classical critical point in $(d+z)$, dimensions, where $z$ is the dynamic critical exponent. 
We shall be primarily concerned with $z=1$, although an example involving $z\ne 1$  is  given below. This extra dimension (imaginary time) is the {\em sine qua non} of the effect of quantum mechanics. 

It seems paradoxical that often the region of classical critical fluctuations is small and tends to zero as the temperature, $T$, tends to zero, while the region of quantum critical fluctuations fans out as $T$ increases.  The reason is that the quantum critical region is determined by the dominant microscopic energy scale in the Hamiltonian, whereas the classical critical region is determined by the ratio of the transition temperature, $T_{c}$, to the same scale raised to a positive power (the dimensionality). Since this ratio is usually small compared to unity  (it is $10^{-5} $ in a conventional superonductor),  the classical critical region is small as well. As a classical critical point is driven to zero by tuning a parameter to reach the QCP, $T_{c}$ itself vanishes as $\xi^{-z}$, provided the $T=0$ spatial correlation length, $\xi$,  is large compared to the lattice spacing, further amplifying the contrast between classical and quantum criticalities. 

How high in temperature does the effect of a QCP persist? Consider perhaps the simplest possible example of a quantum phase transition described by the Hamiltonian of an Ising model in a transverse field \cite{Pfeuty:1970} in one dimension,
\begin{equation}
H=-\sum_{n}\sigma_{1}(n)-\lambda_{1}\sum_{n} \sigma_{3}(n)\sigma_{3}(n+1)
\end{equation}   
where the usual Pauli matrices,  denoted by the $\sigma$s, are situated on a one-dimensional lattice labelled by $n$, of lattice spacing $a$. Here we have scaled out an overall energy scale denoted by $J$. When  $\lambda_{1}\gg1$, the ground state is that of a classical one-dimensional Ising model, which is ordered at zero temperature. The transverse field represented by the Pauli matrix  $\sigma_{1}$ introduces quantum fluctuations and tries to restore the broken $Z_{2}$  symmetry. In the extreme limit $\lambda_{1}\to 0$, the system is a collection of isolated two-level systems and the ground state of each is the symmetric linear superposition of up and down spin states.  It is well known that there is a QCP at the value of the dimensionless coupling constant $\lambda_{1}=1$ , belonging to the universality class of the famous two-dimensional classical Ising model solved by Onsager. The scale $J$ enters as soon as we attempt to calculate the free energy density at finite temperature $T$, which is (setting the Boltzmann constant $k_{B}=1$)
\begin{eqnarray}
f_{0}&=&e_{0}-\frac{T^{2}}{\hbar c}\left[\frac{2aJ\sqrt{\lambda_{1}}}{T}\int_{0}^{\pi/a}\frac{dk}{\pi}\ln \left(1+e^{-\beta J\Lambda_{k}}\right)\right] \nonumber \\
&=&e_{0}-\frac{T^{2}}{\hbar c}\Phi(\lambda_{1},J/T)
\label{eq:f0}
\end{eqnarray}
where $\beta=1/T$, and $\Phi$ is the dimensionless scaled free energy. The quantity $\Lambda_{k}$   denotes the excitation spectrum of the Jordan-Wigner fermions \cite{Pfeuty:1970}:
\begin{equation}
\Lambda_{k}=2\left(1+\lambda_{1}^{2}+2\lambda_{1}\cos ka\right)^{1/2}
\end{equation}
and  $e_{0}$  is the ground state energy density. The quantity  $c=2Ja\sqrt{\lambda_{1}}/\hbar$ is the velocity of the elementary excitations at low energies and is clearly non-universal.

For the validity of QCP at finite temperatures,  the free energy density should satisfy the constraints of finite size scaling \cite{Privman:1984}, namely it should be expressible in terms of a universal scaling function $\Phi_{s}$. In the neighborhood of the QCP, this function  has as its argument the ratio of two fundamental lengths  and approaches a pure number in the limit $T \to 0$, when tuned to the exact critical point. The extent to which $\Phi$ approximates $\Phi_{s}$ defines the quantum critical scaling regime; $\Phi_{s}$ can be extracted from  Eq.~\ref{eq:f0} and satisfies
\begin{equation}
f_{0}-e_{0} \to -\frac{T}{L_{\tau}}\Phi_{s}\left(\left[L_{\tau}/\xi_{\infty}\right]^{1/\nu}\right) ,
\end{equation}
where $L_{\tau}=\hbar c/T$ is the thickness in the imaginary time direction and $\xi_{\infty}\propto 1/|1-\lambda_{1}|^{\nu}$ is the correlation length when $T=0$ , that is, when the thickness $L_{\tau}=\infty$; the exponent $\nu=1$ for the present model. 

If we tune to exact quantum criticality, that is, $\lambda_{1}=1$, then it is known that \cite{Affleck:1986,Bloete:1986} $\Phi_{s}(0)=\pi/12$. Since this is the best chance for quantum criticality to survive at finite temperatures, we test our free energy at this point. The result is shown  in Fig.~\ref{Phi1}. 
It is quite remarkable that scaling holds to a temperature as great as $J/2$. A large overall energy scale helps ensure quantum criticality at high temperatures.

There are always irrelevant operators, however, whose effects must disappear before scaling can be observed. The above Hamiltonian was constructed to have no irrelevant operators. We now turn our attention to a modification of the above Hamiltonian by adding an irrelevant operator involving a three-spin interaction term. This term is generated in the first step of a real space renormalization group procedure \cite{Hirsch:1979}; here we allow its coefficient to be arbitrary. The new Hamiltonian is 
\begin{equation}
H=-\sum_{n}\sigma_{1}(n)\left[1+\lambda_{2}\sigma_{3}(n-1)\sigma_{3}(n+1)\right] - \lambda_{1}\sum \sigma_{3}(n)\sigma_{3}(n+1)
\end{equation}
The only change now is that the transverse field is modified by the nearest-neighbor spins. We have solved this Hamiltonian, once again, by Jordan-Wigner transformation. The fermionic spectrum 
 \begin{equation}
 \Lambda_{k}=2\left(1+\lambda_{1}^{2}+\lambda_{2}^{2}+2\lambda_{1}(1-\lambda_{2})\cos ka -2\lambda_{2}\cos 2 ka\right)^{1/2}
 \end{equation}
vanishes at $ka=\pm\pi$ , when $\lambda_{2}=1-\lambda_{1}$  , and at $k=0$, when  $\lambda_{2}=1+\lambda_{1}$, where $\lambda_{1},\lambda_{2} > 0$, thus defining two critical lines (see Fig.~\ref{Critical}). 

The introduction of the irrelevant operator with the coupling $\lambda_{2}$   does not change the universality class because the correlation length exponent is still given by $\nu=1$ all along the critical lines. It is sufficient to examine the free energy density for $\lambda_{2}=1-\lambda_{1}$ and $0\le\lambda_{1}\le 1$. It is
\begin{eqnarray}
f&=&e_{0}-\frac{T^{2}}{\hbar c}\left[\frac{2aJ|\lambda_{1}-2|}{T}\int_{0}^{\pi/a}\frac{dk}{\pi}\ln \left(1+e^{-\beta J\Lambda_{k}}\right)\right] \nonumber\\
&=&e_{0}-\frac{T^{2}}{\hbar c}\Phi(\lambda_{1},\lambda_{2}=1-\lambda_{1},J/T)
\end{eqnarray}
where $\hbar c = 2Ja\sqrt{4\lambda_{2}+\lambda_{1}(1-\lambda_{2})}=2Ja|\lambda_{1}-2|$ .  In Fig.~\ref{Phi2} we plot the function  $\Phi(\lambda_{1},\lambda_{2}=1-\lambda_{1},J/T)$. The results show that as we move along the critical line starting at  $(\lambda_{2}=0, \lambda_{1}=1)$ and ending at  $(\lambda_{2}=1,\lambda_{1}=0)$, we can dramatically alter the regime of validity of quantum critical scaling at finite temperatures.

\begin{figure}[htb]
\includegraphics[scale=0.9]{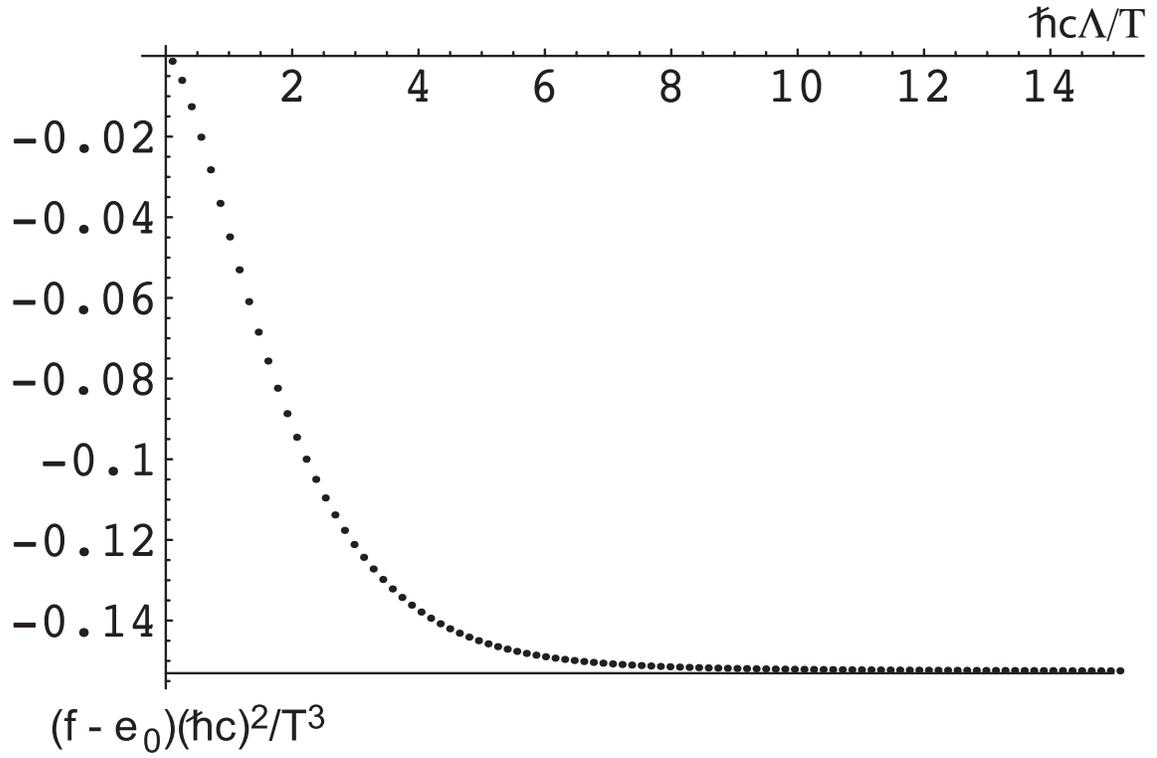}
\caption{\label{onfeplot} The  free energy density minus the ground state  energy density, $(f-e_{0}) (\hbar c)^2/T^3$, of the $O(N)$ quantum non-linear $\sigma$-model in the limit $N \to \infty$ (dotted line).  In the scaling limit ($T \to 0$), this quantity approaches the asymptote $-0.153\cdots$ (solid line).  Quantum critical scaling holds up to $T^* \approx \hbar c \Lambda/8$ with 1\% accuracy.}
\end{figure}

We analyze another model that is important in the study of antiferromagnetism and high-temperature superconductors, the quantum $O(N)$-non linear $\sigma$-model \cite{Chakravarty:1989}.  Taking $d=2$, we consider the partition function
\begin{equation}
Z= \int \mathcal{D} \mathbf{n} \; \delta ( \mathbf{n}^2 -1) \, e^{-{\frac{1}{2g_1}} \int_{0}^{\hbar c \beta} d x_0 \int d^2 x ( \partial_\mu \mathbf{n} )^2}
\label{nlsm}
\end{equation}
where $\mu=(0,1,2)$, and $\mathbf{n}$ is a $N$-component unit vector.  We obtain an exact expression for the free energy in the limit $N\to \infty$, as  in Ref.~\cite{Chubukov:1994}.  The only difference  is that we compute the free energy numerically without assuming $T/\hbar c \ll \Lambda$  and follow the same procedure as before:  tune to the QCP, $g_{1c}$, and look for a breakdown of finite-size scaling.  As shown in Fig. ~\ref{onfeplot}, scaling is valid up to a temperature $T^* \approx \hbar c \Lambda/8$.  Using spin wave theory and an appropriate lattice regularization \cite{Chakravarty:1989}, we obtain $T^* \approx J/2$, where $J$ is the antiferromagnetic exchange constant of the Heisenberg model, which is of order 1000 K in cuprates.

Does $T^*$ change in the presence of additional irrelevant couplings?  We add the simplest irrelevant perturbation that permits an exact solution in the $N \rightarrow  \infty$ limit, namely $g_2 \, \mathbf{n} \cdot \partial_\mu^4 \, \mathbf{n}$.    Using Eq.~\ref{saddlept} given in the Methods section, we have estimated $T^*$ at several critical points $(\tilde{g}_{1c},g_{2c})$ and found that it changes very little as we move along the critical surface.  Hence the irrelevant coupling $g_2$ seems to be truly innocuous, in contrast to the parameter $\lambda_2$ in our previous model.  The difference is not so surprising:  in the latter case we added a perturbation that altered the low-energy excitation spectrum---not only by renormalizing the velocity, but, more importantly, by introducing a secondary minimum.  The term we added here, on the other hand, has no effect whatsoever in the limit $\omega,k \rightarrow 0$, leaving the original low-energy spectrum intact. A different approach to the present model has also suggested  $T^* \approx J/2$ \cite{Chubukov:1994b}. Our analysis of the leading irrelevant operator and our consideration of the free energy have strengthened this result.

In the cuprate superconductors, quantum criticality leads to a powerful and testable prediction relating the transition temperature, $T_{c}$, and the $T=0$ superfluid density, $\rho_{s}(0)$. Consider first the case of a pure system with  $(d+z)$-dimensional QCPs at the edges of the superconducting dome at dopings $x\equiv x_c$. In general the universality classes of these transitions are of course different.  Scaling \cite{Sachdev:1999,Fisher:1989} leads to $T_{c}\propto |x-x_{c}|^{z\nu_{d+z}}$ and $\rho_{s}(0)\propto |x-x_{c}|^{2\beta_{d+z}-\eta_{d+z}\nu_{d+z}}$ for $x$ close to $x_c$ (with standard definitions of critical exponents).  Therefore,  $T_{c}\propto\rho_{s}(0)^y$, where $y$ is determined by the universality class of the transition in question.  If $(d+z)$ is such that  hyperscaling holds (below the upper critical dimension $d_{u}$), then  $y=z/(d+z-2)$, independent of $\nu_{d+z}$.   Moreover, if $d=2$, even $z$ cancels out from the formula, resulting in the superuniversal result $y=1$. Remarkably, this is also the Uemura plot, further elaborated by Emery and Kivelson \cite{Emery:1995}.  If $(d+z) > d_{u}$, the Gaussian fixed point is stable,  and $y=z\nu_{d+z}/(2\beta_{d+z}-\eta_{d+z}\nu_{d+z})=z/2$ because $\nu_{d+z}=\beta_{d+z}=1/2$, $\eta_{d+z}=0$.  

Since the coupling to nodal fermions is irrelevant \cite{Balents:1998}, the QCP in the underdoped regime  belongs  to the XY universality class, with $z=1$ and $d_{u}=4$.  Asymptotically close to the QCP,  the transition must  correspond to $d=3$, but  there is a regime further away in which the layers are effectively decoupled due to weak coupling between them, hence the spatial dimensionality can be set to 2. The details of this crossover depend on various microscopic parameters that are beyond the scope of a scaling theory and must be established empirically. Thus, there is a regime in which Uemura scaling holds. But the true criticality is given by $y=1/2$, modulo weak logarithmic corrections.  In contrast, the QCP in the overdoped regime reflects a  superconductor-to-metal transition, suggesting $z>2$ and $y>1$.  

Strong disorder and macroscopic inhomogeneity can of course destroy the QCPs and therefore scaling. Of more interest to us is weak disorder that modulates  $x_{c}$ (analogous to modulation of $T_{c}$ in a classical phase transition), which may not change the scaling arguments. The reason is as follows: a simple extension of the famous Harris criterion to the $T=0$ quantum problem implies that the exponents of the pure system remain unchanged if $\nu_{d+z}>2/d$. If this inequality is violated, there are two possiblities: (1) the system flows to a new fixed point, in which case the scaling arguments remain valid but  with new exponents; (2) the transition is seriously rounded and the QCP itself  is destroyed. As in any problem involving disorder, {\em a priori} it is difficult to know which of the two possibilities holds. The robustness of the scaling argument is quite striking, however.   When $(d+z)<d_{u}$, $\nu_{d+z}$ cancels out, and even $z$ cancels out in $d=2$; when $(d+z)>d_{u}$, $y$ is simply $z/2$. 

Experiments in high quality crystals \cite{Liang:2005} do in fact yield a value of $y=0.61$, substantially less than unity, in the extreme underdoped regime.  Moreover, based on our analysis, a plot of $T_c$ vs. $\rho_{s}(0)$ should rise more steeply in the underdoped regime than in the overdoped regime; a number of experiments appear to support this assertion at a qualitative level \cite{Niedermayer:1993,Uemura:1993,Bernhard:1995} (an effect known in the literature as the ``boomerang'' effect), although more data are necessary, especially because there are other experiments that do not fit into this picture \cite{Panagopoulos:2003}.

Clearly the role of irrelevant operators is a limiting factor \cite{Belitz:2004} in observing scaling, but, as the term suggests, it is difficult to know ahead of time the set of all such operators.   In some cases unexpected subtleties can arise at very low temperatures, as in a recent experiment involving  $\mathrm {LiHoF_{4}}$ \cite{Ronnow:2005}. In this case, the QCP dictated by the electronic spins is forestalled by the hyperfine coupling of electronic and nuclear spins.  The true QCP is the one at which both the electronic and nuclear spins are entangled.  This entanglement shows up as a transfer of intensity from the magnetic excitation of electronic origin to soft modes of the coupled electronic and nuclear origin in the $10 \mu$eV range. The result is an additional  quasielastic peak at the measured temperatures due to thermal decoherence.  In systems like the cuprates, where the dominant microscopic energy scale (the exchange constant, for example)  is very large, such effects are not of much consequence, and the QCP determined by the larger scale provides the correct picture. Another aspect of quantum criticality is the quantum critical fan \cite{Hertz:1976,Chakravarty:1989,Sachdev:1999} that is determined by the crossover temperatures, $T_{x}$, generically given by $T_{x}=A|x-x_{c}|^{z\nu_{d+z}}$.  The amplitude $A$ controls the extent of the fan, but decreasing the product $z\nu_{d+z}$ clearly narrows it, hence the  scaling regime.

The extended transverse-field Ising model appears to be an interesting example, where the QCP of higher symmetry crosses over to a QCP of lower symmetry at low temperatures (see the Methods section). Whether such a mechanism is effective in a real system like the anomalous normal state of the cuprates is presently not known \cite{Anderson:2002}. Can quantum criticality explain linear-in-$T$ resistivity of optimally doped high-$T_{c}$ superconductors to temperatures as high as 1000 K? It has been argued that this is unlikely \cite{Phillips:2004}. Nonetheless, we have shown that simple considerations of quantum criticality at the edges of the superconducting dome can lead to very powerful experimental consequences for cuprates that can be precisely tested.

\subsection{Methods}
\subsection{The inverse correlation length of the $O(N)$ non-linear $\sigma$-model with an  irrelevant operator involving higher order gradient}
The inverse correlation length $m(\tilde{g}_1,g_2,t)$ is determined by the modified saddle point equation
\begin{equation}
1=\tilde{g}_1 \int_{1>4g_2 \varepsilon_k^2} {\frac{d^2 k}{(2 \pi)^2}} \Psi_1(k,g_2,m,t) +\tilde{g}_1 \int_{1<4g_2 \varepsilon_k^2} {\frac{d^2 k}{(2 \pi)^2}} \Psi_2(k,g_2,m,t)
\label{saddlept}
\end{equation}
where $\varepsilon_k^2=k^2+g_2 k^4+m^2$, $t=T/\hbar c$, and $\tilde{g}_1=(N-1)g_1$.  The integrands $\Psi_1$ and $\Psi_2$ are given by
\begin{equation}\begin{split}
&\Psi_1(k,g_2,m,t)={\frac{\omega_+\coth(\omega_-/2t)- \omega_-\coth(\omega_+/2t)}{2g_2\omega_-\omega_+(\omega_+^2-\omega_-^2)}}\\
&\Psi_2(k,g_2,m,t)={\frac{1}{4g_2\nu_+\nu_-(\nu_+^2+\nu_-^2)}} \times \\
&\left[ {\frac{\nu_+\cot(\nu_-/2t)\mbox{csch}^2(\nu_+/2t)+\nu_-\coth(\nu_+/2t)\csc^2(\nu_-/2t)}{\coth^2(\nu_+/2t)+\cot^2(\nu_-/2t)}} \right] \nonumber
\end{split}\end{equation}
with $\omega_{\pm}=[(1\pm\sqrt{1-4g_2\varepsilon_k^2})/2g_2]^{1/2}$ and $\nu_{\pm}=[(2\sqrt{g_2}\varepsilon_k\pm1)/4g_2]^{1/2}$.

\subsection{Duality transformation}
\begin{figure}[htbp]
\begin{center}
\includegraphics[scale=0.9]{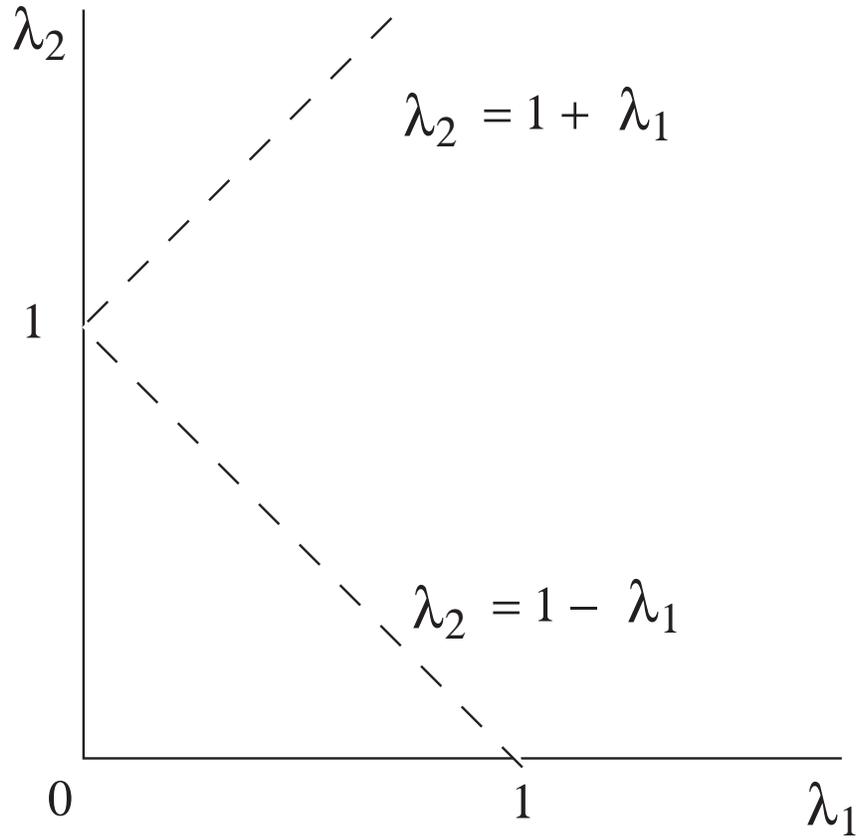}
\caption{The critical lines of the Ising model in a transverse field with three spin interaction. There is a region of parameter space near $\lambda_{1}=0$  and $\lambda_{2}=1$   in which the system will be close to both critical lines, meaning it has low energy excitations at both   $k=0$ and $ka=\pm\pi$, with gaps $\Delta_{0}=2J|\lambda_{1}-(\lambda_{2}-1)|$ and  $\Delta_{\pi}=2J|\lambda_{1}+(\lambda_{2}-1)|$ respectively.  $\Delta_{0}$ collapses along $\lambda_{2} =1+\lambda_{1}$ and $\Delta_{\pi}$ collapses along $\lambda_{2} =1-\lambda_{1}$. In either case, the critical exponent is given by $\nu=1$. Under the duality transformation described in the Methods section, the multicritical point $(\lambda_{1}=0,\lambda_{2}=1)$ maps onto the isotropic fixed point of the quantum XY model in a transverse field.}
\label{Critical}
\end{center}
\end{figure}

\begin{figure}[htbp]
\begin{center}
\includegraphics[scale=0.85]{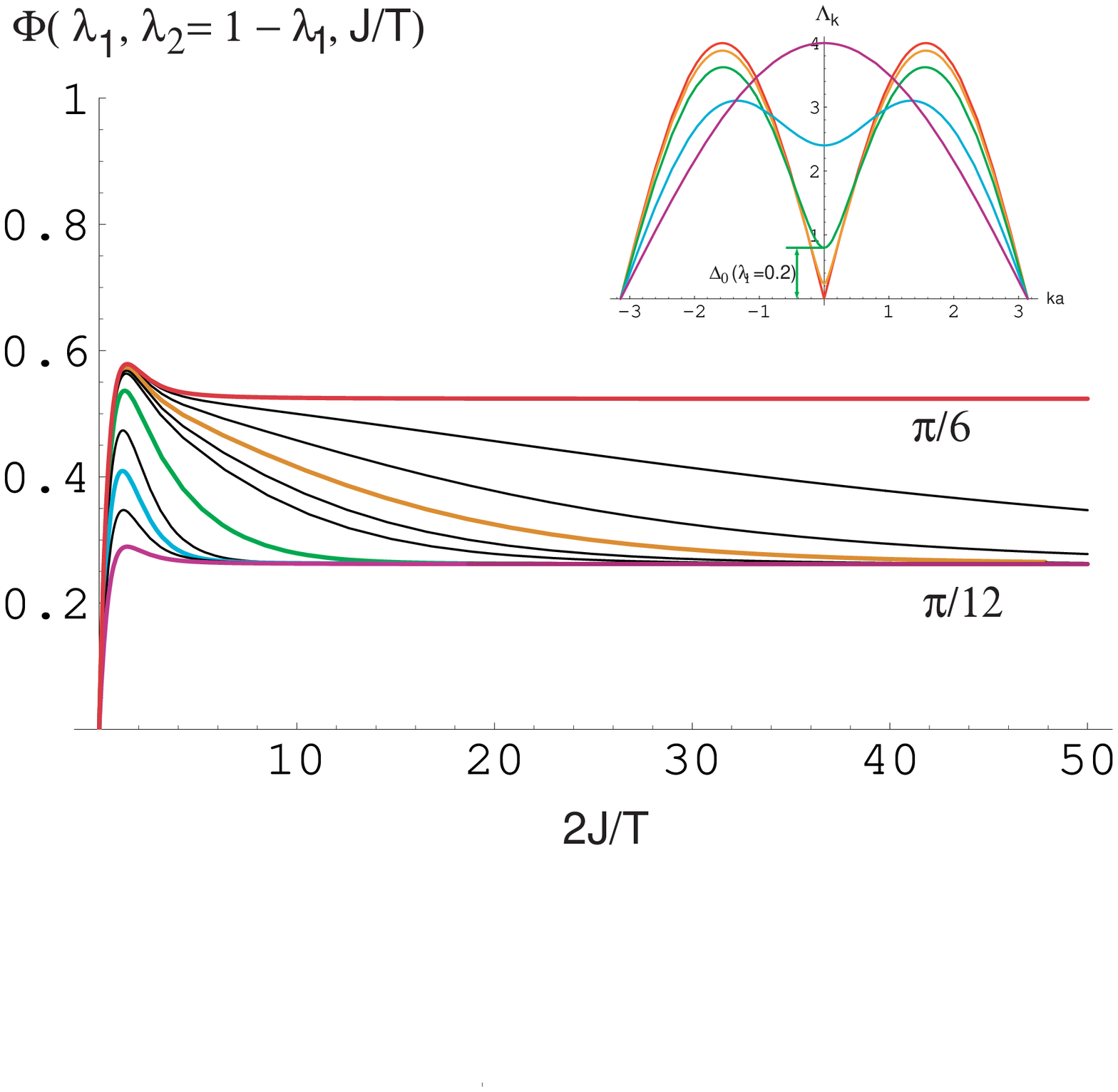}
\caption{The quantity $\Phi(\lambda_{1},\lambda_{2}=1-\lambda_{1},J/T)$  is plotted as a function of  $2J/T$ for different values of $\lambda_{1}$ while maintaining criticality, that is, $\Delta_{\pi}=0$. From top to bottom, the values of   $\lambda_{1}$ are 0, 0.02, 0.04, 0.06, 0.08, 0.1, 0.2, 0.4, 0.6, 0.8, 1.0. Five of these curves are shown in color, with the corresponding excitation spectra displayed in the inset.  In every case, the gapless excitations near $ka=\pm \pi$ are responsible for scaling; but as we approach the multicritical point $(\lambda_2 = 1,\lambda_1 =0)$, a new set of low energy modes appears at $k=0$, with the gap $\Delta_{0}/J = 2|\lambda_1 - (\lambda_2 -1)|=4\lambda_1$.  $\Phi$ does not approach $\Phi_s =\pi/12$ until $T\ll \Delta_0$, which requires arbitrarily low temperatures as $\Delta_0 \to 0$.  Precisely at the point $(\lambda_2 = 1,\lambda_1 =0)$, $\Delta_0$ collapses and the number of critical modes is doubled, giving rise to a new asymptote $\Phi_s=\pi/6$.}
\label{Phi2}
\end{center}
\end{figure}

By a duality transformation \cite{Fradkin:1978} where the dual quantum spins ($\tau_{i}(n)$) satisfying the Pauli spin algebra are located at the centers of the bonds of the original lattice, we can rewrite our extended model (Eq.~5) as 
\begin{equation}
H=-\sum_n \left[\lambda_1 \tau_1(n)+ \tau_3(n) \tau_3(n+1) - \lambda_2 \tau_2(n) \tau_2(n+1)\right],
\end{equation}	
which has been recently studied in the context of entanglement \cite{Wei:2005}. The point ($\lambda_{1}=0, \lambda_{2}=1$) is the unstable isotropic XY fixed point; note that ($\lambda_{1}=0, \lambda_{2}=-1$) has the same excitation spectrum. For temperatures $T\gg T^{*}\propto \Delta_{0}$ (see, Figs.~\ref{Critical} and \ref{Phi2}), the behavior is controlled by this unstable fixed point. An integrable perturbation induces a flow to the Ising fixed point at lower temperatures. In the language of conformal field theory, this crossover is described by Zamolodchikov's ``$c$-function'', which takes one from a theory of conformal charge $c=1$ to that of charge $c=1/2$ \cite{Castro-Neto:1993}.

\vspace{0.5cm}

We thank E. Fradkin,  S. Sachdev, S. L. Sondhi,  and A. P. Young for helpful comments on the manuscript and  H. M. R{\o}nnow and D. Bonn for the prepublication copies of Refs.~ \cite{Ronnow:2005} and \cite{Liang:2005}. This work was supported by the NSF under grant:  DMR-0411931.

\newpage

\begin{figure}[htbp]
\begin{center}
\includegraphics[scale=0.9]{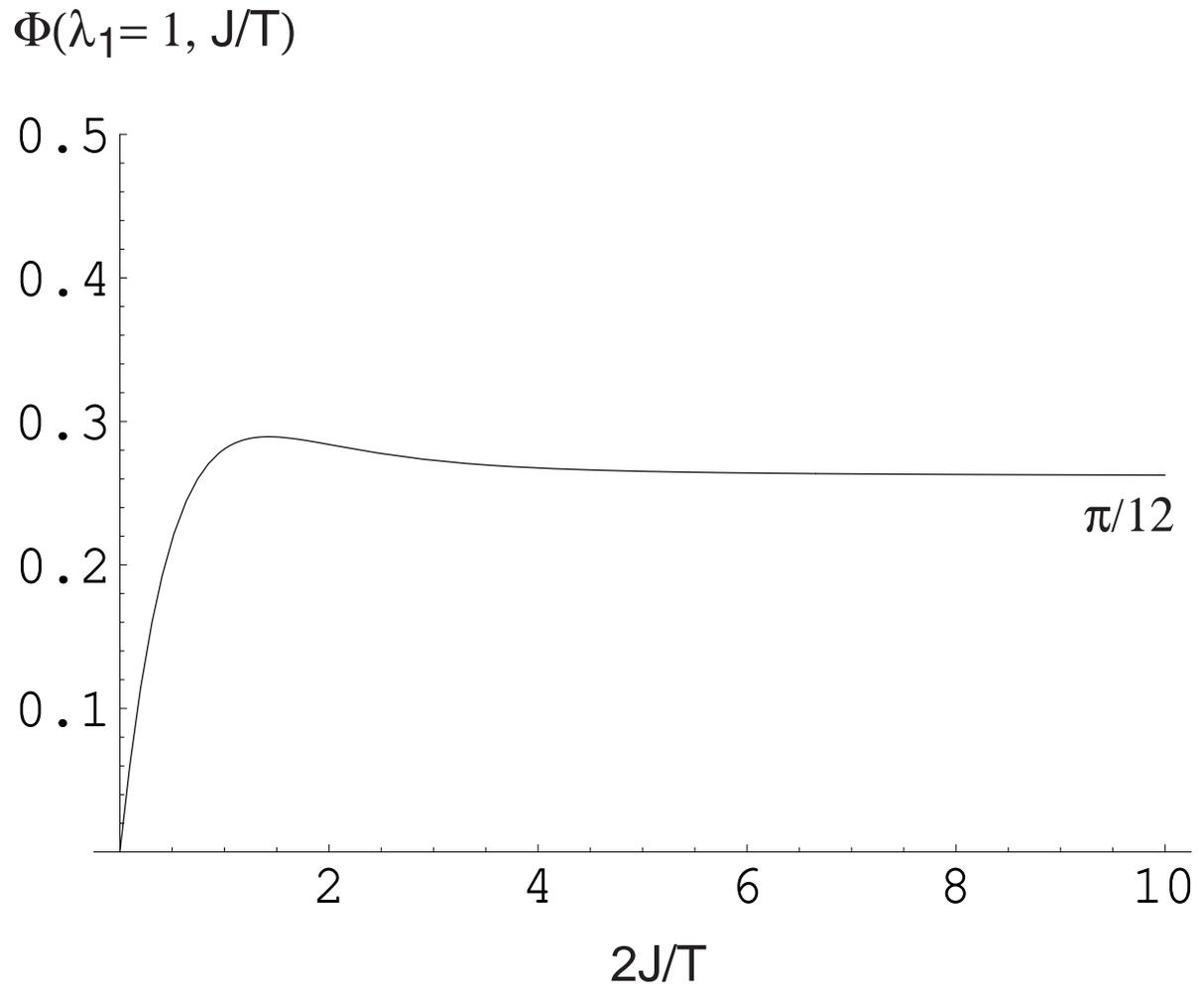}
\caption{The scaled free energy $\Phi(\lambda_{1}=1, J/T)$ plotted as a function of $2J/T$. The   asymptote as $T\to 0$ should be the universal number $\pi/12$, if quantum critical scaling holds at finite temperatures. It clearly does so with sufficient accuracy for temperatures as large as $J/2$.}
\label{fPhi1}
\end{center}
\end{figure}

\end{document}